# High-Energy Threshold Reaction Rates on 0.8 GeV Proton-Irradiated Thick W and W-Na Targets

YU.E. TITARENKO[*], V.F. BATYAEV, E.I. KARPIKHIN, V.M. ZHIVUN, A.B. KOLDOBSKY,
R.D. MULAMBETOV, S.V. MULAMBETOVA, S.L. ZAITSEV, S.G. MASHNIK [1], R.E. PRAEL [1]

*Institute for Theoretical and Experimental Physics (ITEP), B.Cheremushkinskaya 25, 117259 Moscow, Russia*
[1] *Los Alamos National Laboratory, Los Alamos, NM 87545, USA*

**Abstract** Results are presented of measuring the threshold activation reaction rates in $^{12}$C, $^{19}$F, $^{27}$Al, $^{59}$Co, $^{63}$Cu, $^{65}$Cu, $^{64}$Zn, $^{93}$Nb, $^{115}$In, $^{169}$Tm, $^{181}$Ta, $^{197}$Au, and $^{209}$Bi experimental samples placed along the axis inside and outside the 0.8 GeV proton-irradiated 92-cm thick W-Na and 4-cm thick W targets. 158 reactions of up to ~0.5 GeV thresholds have been measured in 123 activation samples for W-Na target, and 157 reactions in 36 activation samples for W target. The reaction rates were determined using the γ-spectrometry method. In total, more than 1000 values of activation reactions were determined in the experiments. In both cases the measured reaction rates were compared with the LAHET code simulated rates and using several nuclear databases for the respective excitation functions, namely, ENDF/B6 for cross section of neutrons at energies below 20 MeV and MENDL2 together with MENDL2P for cross sections of protons and neutrons of 20 to 100 MeV energies. A general satisfactory agreement between simulated and experimental data has been found. Nevertheless, further studies should be aimed at perfecting the simulation of the production of the secondary protons and high-energy neutrons, especially in the sideward and backward directions with respect to the beam. The results obtained permit some conclusions concerning the reliability of the transport codes and databases used to simulate the ADS with Na-cooled W targets. The high-energy threshold excitation functions to be used in the activation-based unfolding of neutron spectra inside the ADS can also be inferred from the results.

## I. INTRODUCTION

The pending researches with the pilot Accelerator Driven Systems (ADS) require reliable nuclear data. One of the possible fields of the data application is the activation-based unfolding of the high-energy (up to ~1 GeV) neutron spectra inside the ADS target and the near-target blanket zone. The high-energy "tail" in the ADS spectra triggers the high-energy threshold (>10 MeV) reactions, which are not studied in detail in the conventional reactor researches. Therefore, the excitation functions that may be derived from high-energy calculations, or retrieved from the available databases require insistently a reliable verification via testing experiments with the high-energy proton-irradiated target micromodels.

The ITEP U-10 proton synchrotron was used to realize a run of experiments to study the threshold activation reaction rates inside and outside thick W and W-Na targets.

## II. EXPERIMENT

The W-Na target was assembled of alternating cylindrical W and Na discs (see Fig. 1A). The displayed succession of the disks was selected to obtain the maximum attainable uniform neutron distribution along the target. The W discs are the 150-mm diameter full-metal structures of 40-mm (7 pieces) and 20-mm (5 pieces) thicknesses. The Na discs (13 pieces) are the 150-mm diameter, 40-mm thick thin-walled cylindrical containers made of 0.5-mm thick stainless sheet steel, filled with metallic Na, and soldered tight. The experimental activation samples are 10.5-mm or 10.0-mm diameter circular discs cut of metal foils of the certified composition. During the irradiation runs, the samples were placed on the W disc surfaces and in their seats milled in the rulers (see Fig. 2). In the experiments, the activation reaction rates were determined on the $^{209}$Bi, $^{197}$Au, $^{169}$Tm, $^{115}$In, $^{93}$Nb, $^{65}$Cu, $^{64}$Zn, $^{63}$Cu, $^{59}$Co, $^{27}$Al, $^{19}$F, and $^{12}$C nuclides. Besides, the $^{22}$Na production rate in the Na discs was determined.

The W target is a single W disk of 150-mm diameter and 40-mm thickness. The same experimental activation samples, except $^{169}$Tm and $^{181}$Ta, were placed on the W-disk, as shown in Fig. 1B.

---

[*] *Corresponding author: E-Mail Yury.Titarenko@itep.ru*

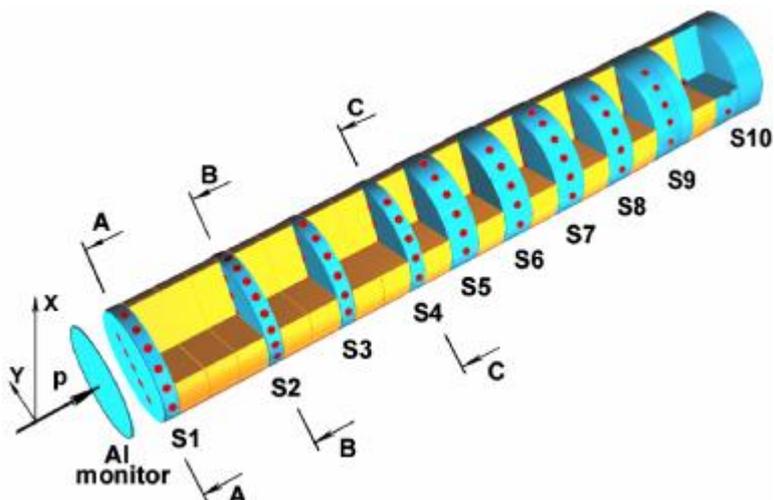

Fig 1A. The W-Na target assembly view.

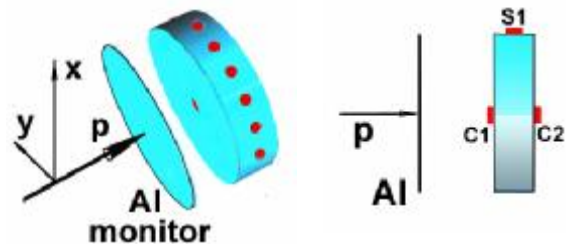

Fig. 1B. A schematic of the arrangement and irradiation of the samples of a single W disc.

Both targets were irradiated with the ITEP synchrotron-extracted 0.8 GeV proton beam. The W-Na and W irradiations lasted for 10 h and 2 h, respectively. The proton number on the targets was found to be $(6.5\pm0.4)*10^{14}$ for W-Na target and $(6.0\pm0.4)*10^{13}$ for W target. The irradiated activation samples were $\gamma$-spectrometered using the Ge and Ge-Li detector-based high-resolution spectrometers. The details of target composition, experimental layout, irradiation parameters, measuring technique and reaction rates determining can be found in [1, 2].

In total, 979 and 296 values of the reaction rates have been determined in the W-Na and W experiments, respectively (see Table 1). Comparison was made between the sets of the high-threshold reaction rates in the case of identical arrangement of the respective samples in the W and W-Na targets. Fig. 3 shows the reaction rates in Bi and In measured at point S1 in both experiments. It should be noted that the excess of the reaction rates in W over W-Na rises as the reaction threshold increases. In the case of the high-threshold reactions (with $E_{th}>200$ MeV), the experimental W – to - W-Na reaction rate ratio reaches a few unities.

Table 1. List of the reaction rates measured.

| Nuclide | The products for the reaction rates measured | Number of values W-Na | W |
|---|---|---|---|
| $^{209}$Bi | $^{207}$Po, $^{206}$Po, $^{206}$Bi(i,c), $^{205}$Bi, $^{204}$Bi, $^{203}$Bi, $^{203}$Pb(i,c), $^{201}$Pb, $^{200}$Pb, $^{202}$Tl, $^{200}$Tl(i,c), $^{191}$Pt, $^{96}$Tc, $^{96}$Nb, $^{82}$Br | 165 | 14 |
| $^{197}$Au | $^{198}$Au, $^{196}$Au, $^{194}$Au, $^{191}$Pt | 33 | 6 |
| $^{181}$Ta | $^{182}$Ta, $^{178m}$Ta, $^{176}$Ta, $^{175}$Ta, $^{173}$Ta, $^{180m}$Hf, $^{175}$Hf, $^{173}$Hf, $^{170}$Hf, $^{172}$Lu, $^{171}$Lu, $^{169}$Lu, $^{167}$Tm, $^{165}$Tm, $^{161}$Er | 112 | - |
| $^{169}$Tm | $^{166}$Yb, $^{168}$Tm, $^{167}$Tm, $^{166}$Tm(i,c), $^{165}$Tm, $^{163}$Tm, $^{161}$Er, $^{160}$Er, $^{160m}$Ho(i,c), $^{157}$Dy, $^{155}$Dy, $^{153}$Dy, $^{152}$Dy, $^{153}$Tb, $^{152}$Tb, $^{151}$Tb, $^{150}$Tb | 19 | - |
| $^{115}$In | $^{113}$Sn, $^{116}$In, $^{115m}$In(i,c) $^{114m}$In, $^{113m}$In(i,c), $^{111}$In, $^{110}$In, $^{109}$In, $^{115}$Cd, $^{111}$Ag, $^{110m}$Ag, $^{106m}$Ag, $^{105}$Ag, $^{101}$Pd, $^{100}$Pd, $^{101}$Rh, $^{100}$Rh(i,c), $^{99}$Rh, $^{97}$Ru, $^{96}$Tc, $^{95}$Tc, $^{90}$Mo, $^{99}$Nb(i,c), $^{89}$Zr, $^{87}$Y | 207 | 119 |
| $^{93}$Nb | $^{90}$Nb, $^{89}$Zr, $^{86}$Zr, $^{90}$Y, $^{87m}$Y, $^{87}$Y, $^{86}$Y(i,c) | 8 | 21 |
| $^{65}$Cu | $^{64}$Cu, $^{61}$Cu, $^{65}$Ni, $^{58}$Co, $^{57}$Co, $^{56}$Co, $^{55}$Co, $^{56}$Mn, $^{52}$Mn, $^{44m}$Sc, $^{44}$Sc(i,m+g) | 12 | 9 |
| $^{64}$Zn | $^{65}$Zn, $^{63}$Zn, $^{62}$Zn, $^{64}$Cu, $^{61}$Cu, $^{60}$Cu, $^{57}$Ni, $^{58}$Co, $^{57}$Co, $^{56}$Co, $^{55}$Co, $^{56}$Mn, $^{52}$Mn, $^{48}$Cr, $^{44m}$Sc, $^{44}$Sc(i,m+g) | 17 | 18 |
| $^{63}$Cu | $^{61}$Cu, $^{58}$Co, $^{55}$Co, $^{52}$Mn | 4 | 13 |
| $^{59}$Co | $^{57}$Ni, $^{58}$Co(i,m+g), $^{58m}$Co, $^{57}$Co, $^{56}$Co, $^{55}$Co, $^{59}$Fe, $^{52}$Fe. $^{56}$Mn, $^{52}$Mn, $^{51}$Cr, $^{48}$Cr, $^{48}$V, $^{48}$Sc, $^{47}$Sc(I,c), $^{46}$Sc, $^{44m}$Sc, $^{44}$Sc(i,m+g), $^{43}$Sc, $^{47}$Ca, $^{43}$K, $^{24}$Na, $^{7}$Be | 222 | 73 |
| $^{27}$Al | $^{27}$Mg, $^{24}$Na, $^{22}$Na, $^{7}$Be | 186 | 21 |
| $^{19}$F | $^{18}$F | 1 | 1 |
| $^{12}$C | $^{11}$C | 1 | 1 |
| | **Total** | **979** | **296** |

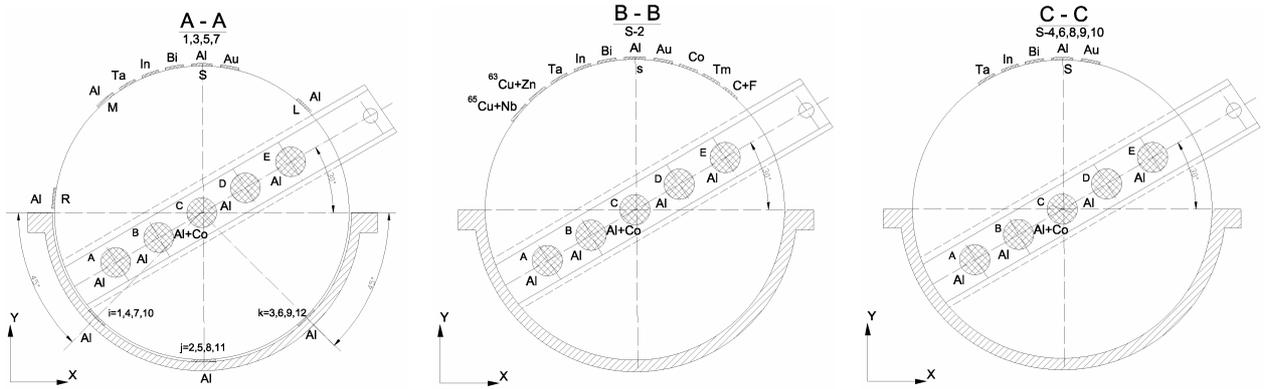

Fig. 2. Arrangement of experimental samples on W discs. The A-A sectional view shows discs Nos. 1, 3, 5, and 7; the BB view shows disc No. 2; the C-C view shows discs Nos. 6, 8, 9, and 10.

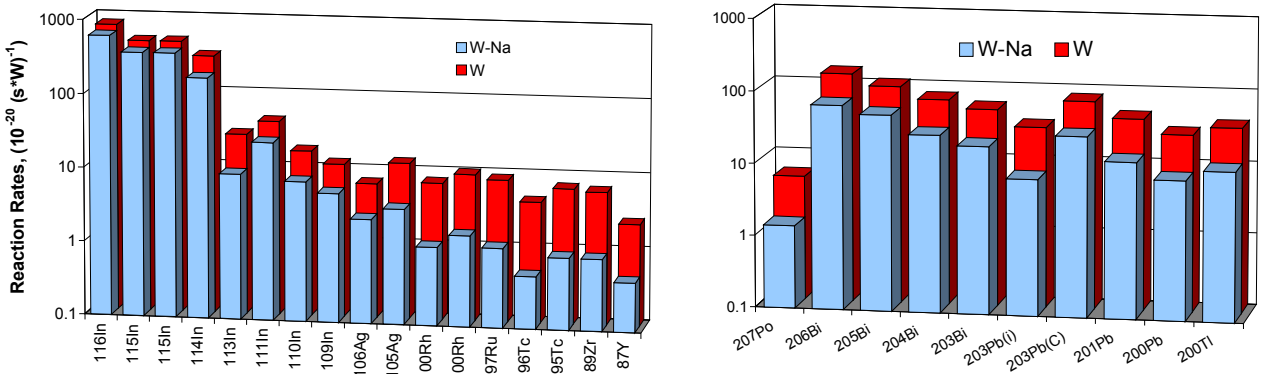

Fig. 3. The reaction rates in the $^{115}$In (left panel) and $^{209}$Bi (right panel) samples measured at point S1 for the W and W-Na targets

### III. COMPUTATIONAL SIMULATION OF THE MEASURED REACTION RATES

The reaction rates were simulated with the use of the LAHET Code System[3)] including the LAHET and HMCNP codes. The codes simulated the neutron and proton spectra in the experimental sample locations (see Fig. 4).

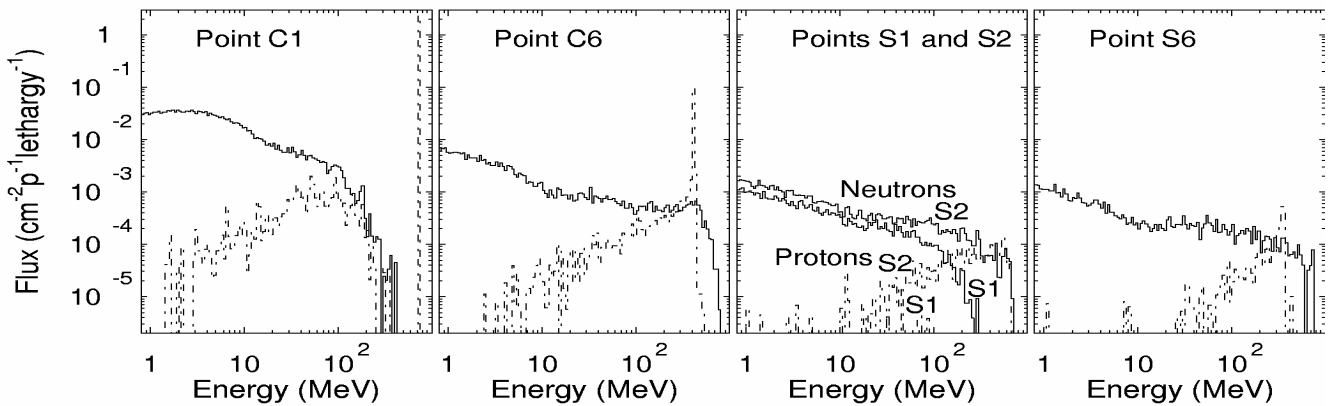

Fig. 4. The calculated neutron and proton fluxes at the selected points inside and outside the W-Na target. The solid and dashed lines show the neutron and proton fluxes, respectively. Points C1 and C6 denote the center of the first and sixth W discs, S1, S2, and S6 - the surfaces of first, second and sixth W discs.

Then, the simulated reaction rates may then be determined via the integral multiplication of the calculated spectra by the respective reaction cross sections: $R_x = R_{n,x} + R_{p,x} = \sum_{i=n,p} \int \sigma_{i,x}(E) \cdot \varphi_i(E) dE$. The required proton and neutron excitation functions $\sigma_{i,x}(E)$ were accumulated using:

- the MENDL2[4] and MENDL2p[5] databases, which give the neutron and proton reaction excitation functions up to 100 MeV and 200 MeV, respectively;
- the LAHET code simulated excitation functions from 100 MeV to 800 MeV;
- the available experimental data on the production cross sections of the respective reaction products (the EXFOR database).

Basing on the above dataset for a particular reaction, we can select the excitation function that would be consistent with all the data and agree satisfactorily with the experimental reaction rates at all measurement points.

The following two groups of reactions, whose cross sections are determined by different algorithms, can tentatively be singled out.

1. The reactions supported by sufficient experimental proton cross section data throughout the necessary energy range. The MENDL and LAHET excitation functions are, then, normalized and joined to each other, so that they would optimally describe the set of experimental points. In this case, the same normalization parameters were also applied to the neutron excitation functions that are not supported by any experimental points.

2. The reactions supported by meager, or even none of, experimental data. In this case, where the comparison between experimental and calculated reaction rates shows systematic deviations of calculations from experiment, LAHET and MENDL were used to normalize the excitation functions in such a manner that the deviations would be minimized.

Our earlier work[1] describes the reaction rates simulation algorithm in details.

## IV. COMPARISON BETWEEN EXPERIMENT AND CALCULATIONS

Our earlier work[1] has shown that most of the experimental results can be simulated to within a satisfactory accuracy (almost a half of the calculated data differ from experiment by less than 30%). The present work deals only with the reactions, for which a significant disagreement with experiments is observed (Fig. 5).

1. The calculated $^{209}$Bi(p,3n)$^{207}$Po and $^{209}$Bi(p,4n)$^{206}$Po reaction rates in the first and eighth W discs are much underestimated compared with experiment. Since $^{207}$Po and $^{206}$Po cannot be produced in neutron reactions on $^{209}$Bi, we think that the discrepancy has arisen most probably from the fact that the LAHET code simulates rather inadequately the secondary proton production in backward direction in a hadron-nucleus cascade. Fig. 3 demonstrates the extremely low simulated proton flux on the surface of the first W disc (point S1).

2. From Figs. 5 it follows that point S1 is characterized by a significant underestimation of the data calculated for not only the proton-induced reactions proper, but also the high-threshold (~100 MeV and higher) mixed-type reactions. Table 3 is the list of the reactions. The disagreement can be explained by Fig. 3 (the third panel), which shows the neutron and proton spectra at points S1 and S2. The comparison between the neutron spectra at the two points indicates that, up to ~50 MeV, the S2 neutron flux density is about twice the S1 neutron flux density, in a good agreement with the $R_{S1}/R_{S2}$ ratio for the moderate-threshold reactions ($^{27}$Al→ $^{24}$Na from Al, for instance). As neutron energy increases, however, the difference between the S1 and S2 flux densities rises, with the S1 flux vanishing at ~300 MeV. If the S1 neutron flux was actually as shown in Fig 3, the reactions with thresholds above ~200-300 MeV would have never been observed at point S1 (for example, $^{115}$In→$^{87}$Y, a ~400 MeV threshold). The significant experimental rates of the high-threshold reactions indicate that the simulated S1 neutron spectrum above ~100 MeV is much underestimated

Table 3. Comparison between experimental and calculated rates of high-threshold reactions at point S1.

| Reactions | Threshold, MeV | Underestimation factor |
|---|---|---|
| $^{201}$Pb, $^{200}$Pb, $^{200}$Tl(i,c) from $^{209}$Bi; $^{173}$Ta, $^{173}$Hf, $^{167}$Tm from $^{181}$Ta | 50 – 100 | 2 – 5 |
| $^{191}$Pt from $^{209}$Bi; $^{101}$Pd, $^{100}$Rh (i,c) from $^{115}$In | 100 – 200 | 10 – 50 |
| $^{100}$Pd, $^{97}$Ru, $^{96}$Tc, $^{95}$Tc, $^{90}$Nb, $^{87}$Y from $^{115}$In | 200 and higher | 100 and higher |

The above mentioned difference in the reaction rates at point S1 in the W and W-Na experiments made us compare the high-energy part of the neutron spectra at that point. The considerable excess of the above presented reaction rates at point S1 indicates a significant difference in the high-energy parts of the neutron spectra. A detailed comparison between the spectra is shown in Fig. 6 as the ratio of the spectra at point S1 and the ratio of experimental reaction rates. A steady rise of the ratio of the simulated spectra with increasing neutron energy (from 1.6 at $E_n$ = 10 MeV to ~2.5 at $E_n$ = 300-400 MeV) is quite evident. At the same time, the ratio of the experimental reaction rates in the 300-700 MeV range reaches ~8. So, the calculated spectral hardening proved to be, as a minimum, 3 times as low as that observed experimentally.

To clarify the causes of the resultant discrepancies, Fig. 7 presents the calculated neutron spectra at points S1 and S2 for the W-Na target together with the experimental and calculated double-differential neutron production cross sections in W at 0,8 GeV. In the case of the double-differential cross sections, the agreement between calculations and experiment is much better, permitting us to conclude that the LAHET simulation of neutron production in all directions cannot give rise to the above mentioned discrepancies.

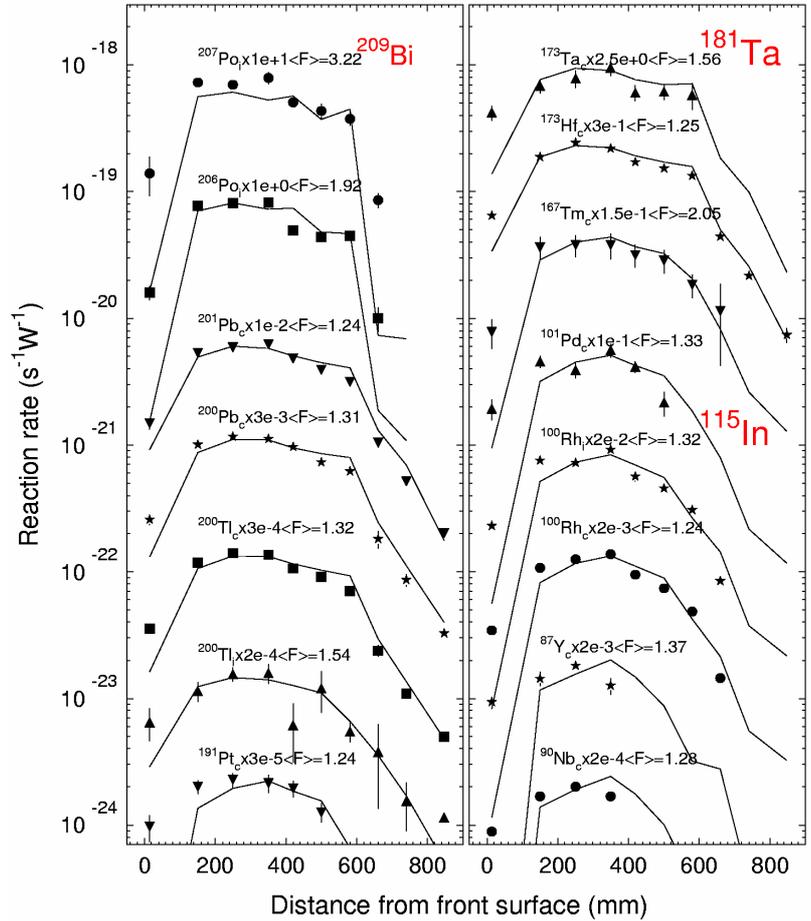

Fig. 5. The calculated and experimental reaction rates. The mean squared deviation factor <F> are presented for each reaction

In our opinion, the LAHET description of the high-energy neutron transport in the W discs is predominantly responsible for the discrepancies. Particularly, this may concern the processes of high-energy neutron scattering from the range of small angles. The noted assumption will be studied in details in further works.

## V. CONCLUSION

The results presented above indicate that, with rare exclusions, the described approaches make it possible to obtain the excitation functions of high-energy threshold reactions that lead to a satisfactory description of the measured reaction rates. The functions can be used to unfold the neutron spectra at different points inside the ADS.

Further studies should nevertheless be aimed at perfecting the simulation of the production of secondary protons and high-energy neutrons, especially in the sideward and backward direction with respect to the beam. This is emphasized by the data on the rates of Po production from Bi and of high-threshold reactions sidewise of the beam entry to the target.

It should be noted that the satisfactory agreement in the reaction rates is somewhat conditional in the case of lacking reliable experimental data on excitation function for Bi, In, Ta, etc. Therefore, additional high-energy experiments are required to measure the production cross section of secondary product nuclei and the reaction rates in the integral experiments with other target types.

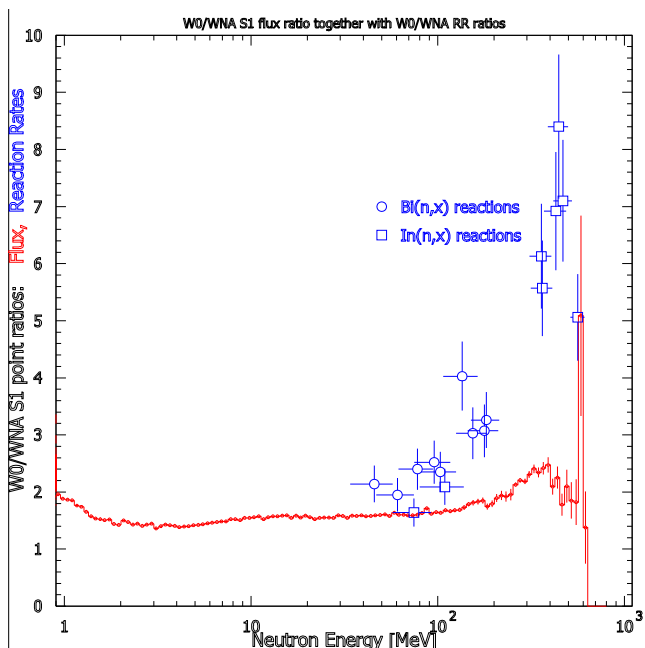
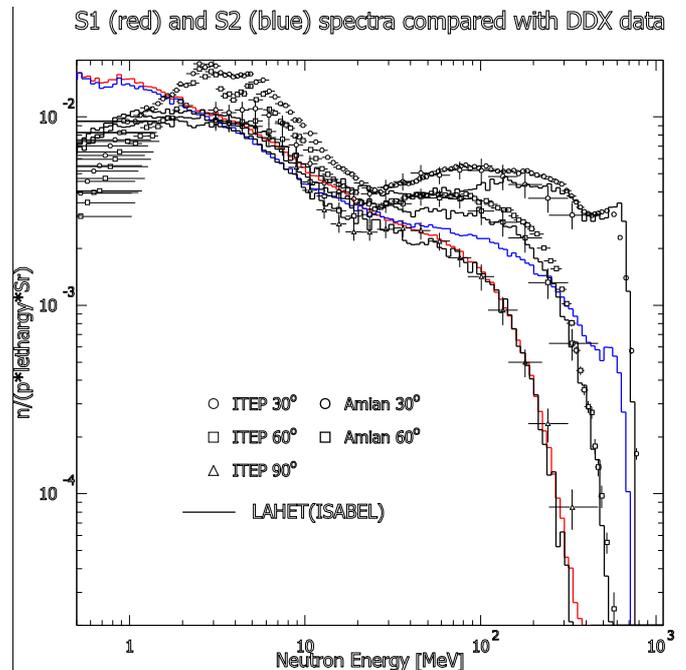

Fig. 6. The LAHET-simulated ratios of the neutron spectrum at point S1 for W target to the spectrum for W-Na target (the red line). The ratios of experimental reaction rates in Bi and In samples at point S1 for W target to the ratios of the same reactions for W-Na target. Plotted as abscissa is the characteristic energy at which a reaction proceeds.

Fig. 7. The simulated neutron spectra at points S1 (the red line) and S2 (the blue line) for W-Na target and the experimental and calculated double-differential cross sections for neutron production in W at angles 30°, 60°, and 90° at proton energy 0.8 GeV. The experimental data are from the works of ITEP[6] and LANL[7].

ACKNOWLEDGEMENT

The authors are indebted to Dr. F.E.Chukreev (Kurchatov Institute) for his assistance in working with the EXFOR database.
The work has been supported by International Science and Technology Center (ISTC) under Project#1145. In addition, the work was partly supported by the U.S. Department of Energy.